\documentclass[journal]{IEEEtran}
\usepackage{graphicx}
\usepackage{epsf}

\usepackage{amsmath}
\usepackage{amssymb}
\usepackage{amsthm}
\usepackage{latexsym}
\usepackage{multirow}
\usepackage{multicol}
\usepackage[table]{xcolor}
\usepackage[hidelinks]{hyperref}
\usepackage{algorithm2e}
\usepackage{subcaption}

\usepackage{mathtools}

\begin{document}
{
\title{\fontsize{24}{28}\selectfont Analysis of Human Exposure to Electromagnetic Fields in 5G Uplink and Downlink}

\author
{
Seungmo Kim, \textit{Member}, \textit{IEEE}


\thanks{S. Kim is with the Department of Electrical and Computer Engineering, Georgia Southern University in Statesboro, GA, USA.}
}

\maketitle
\begin{abstract}
Concern has been widely acknowledged about human health---e.g., heating of the eyes and skin--from exposure to electromagnetic fields (EMF) produced by wireless transmitters. Mobile telecommunications rely on an extensive network of base stations (BSs) and handheld devices that transmit signals via EMF. There is chance of aggravation due to two important changes that will be seen in future cellular networks. First, the number of BSs will remarkably grow with the proliferation of small-cell networks, which will expose humans to EMF more often. Second, highly concentrated EMF beams will be generated by employing larger antenna arrays to overcome faster EMF energy attenuation in higher-frequency bands such as millimeter wave (mmW) spectrum, which will increase damage if the main beam points to the human body. However, the two changes can be exploited as leverages for (i) wider selection of alternative BSs and (ii) more precise beamforming to a desired user equipment (UE) with less EMF leakage to other directions, respectively. Harnessing the two changes, we have been investigating the human health impacts of 5G wireless systems. This extended abstract summarizes our findings thus far.
\end{abstract}

\vspace{0.1 in}


\vspace{0.1 in}

\section{Introduction}\label{sec_intro}
\subsection{Background}
As a means to fulfill the latest skyrocketing bandwidth demand, the fifth-generation wireless (5G) is expected to achieve far higher data rates compared to the previous-generation wireless systems. Recently, however, a serious concern on human health has been raised. The 5G’s high data rate needs a higher signal power at a receiver. Note that an increase in signal power received at a user’s end causes an increase in the electromagnetic energy imposed on the user \cite{appeal}.

Not only that, the present work identifies three technical features adopted in 5G, which potentially increase the concern of human electromagnetic field (EMF) exposure \textit{further}: (i) higher carrier frequencies (e.g., 28, 60, and 70 GHz \cite{jsac17}\cite{verboom20}\cite{kabir19}); (ii) larger number of transmitters due to introduction of small cells \cite{jsac17}; (iii) higher concentration of EMF energy due to adoption of beamforming \cite{jsac17}.

\subsection{Contributions}
The contributions of our research are three-fold:
\begin{itemize}
\item It discusses the human EMF exposure in the downlink as well as the uplink.
\item It work suggests that both specific absorption rate (SAR) and power density (PD) should be used to display human EMF exposure for a wireless system. \item It work presents an explicit comparison of human EMF exposure in 5G to those in the currently deployed wireless standards.
\end{itemize}

\vspace{0.1 in}

\section{Related Work}
\subsection{Health Effect}
Heating is considered as a significant impact since it can cause subsequent effects such as cell damage and protein induction \cite{pall}. The pain detection threshold temperature for human skin is approximately 43$^{o}$C \cite{mine1} and any temperature exceeding it can cause a long-term injury.

\subsection{Regulation Guidelines}
Being aware of the health hazards due to EMF emissions generated in a wireless communication system, the United States (US) Federal Communications Commission (FCC) \cite{fcc} and the International Commission on Non-Ionizing Radiation Protection (ICNIRP) \cite{icnirp} set guidelines on the maximum amount of EMF energy allowed to be introduced in a human body.

\subsection{Measurement Metrics}
PD and SAR are the two most widely accepted metrics to measure the intensity and effects of EMF exposure \cite{mine1}. The FCC suggests PD as a metric measuring the human exposure to EMF generated by devices operating at frequencies higher than 6 GHz \cite{fcc}, whereas a recent study suggested that a guideline defined in PD is not efficient to determine the impacts on health issues especially when devices are operating extremely close to the human body such as in an uplink \cite{mine1}-\cite{mine8}.

\begin{figure*}
\centering
\begin{subfigure}{.495\textwidth}
\centering
\includegraphics[width=\linewidth]{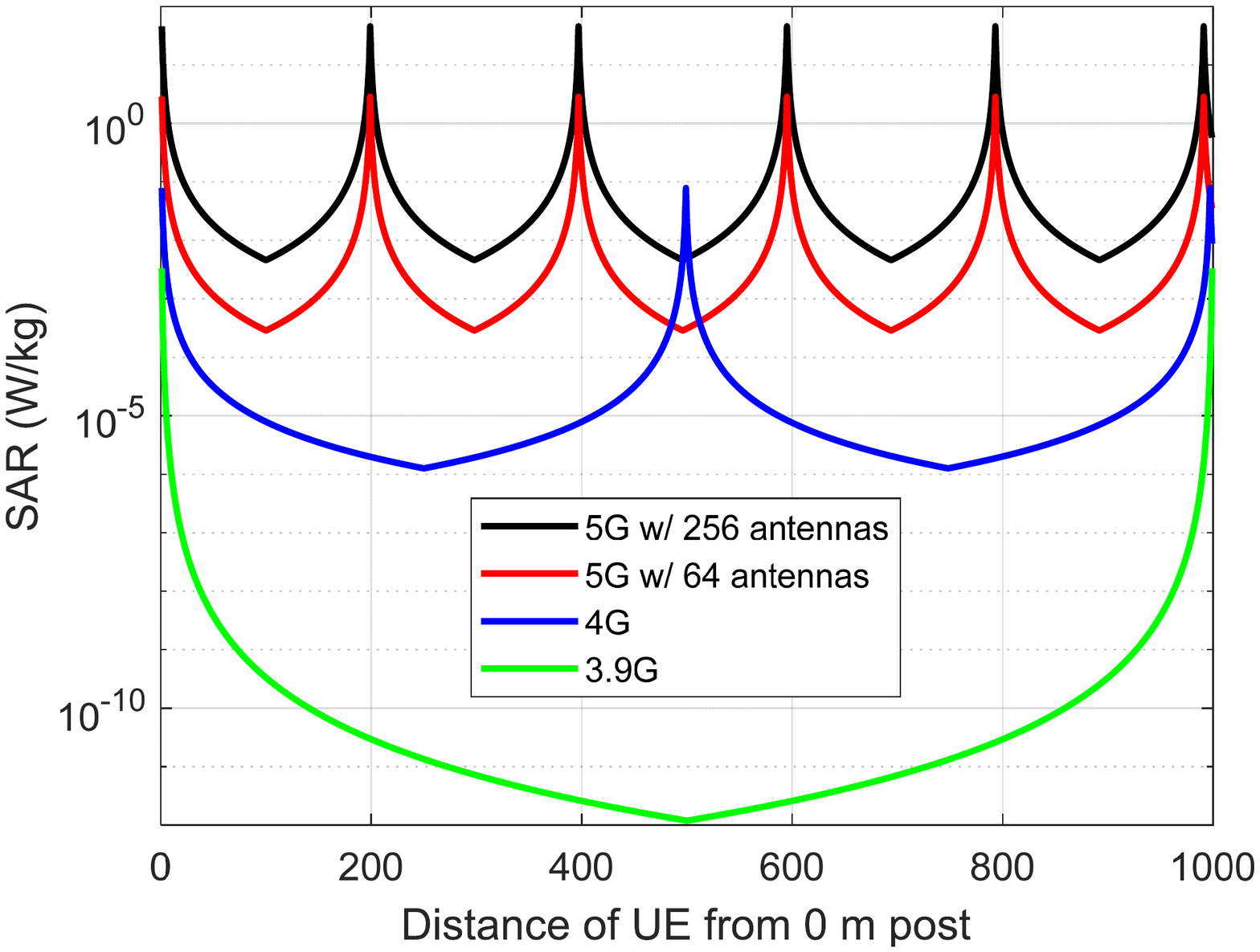}
\caption{Downlink}
\label{fig_downlink}
\end{subfigure}
\hfill
\begin{subfigure}{.495\textwidth}
\centering
\includegraphics[width=\linewidth]{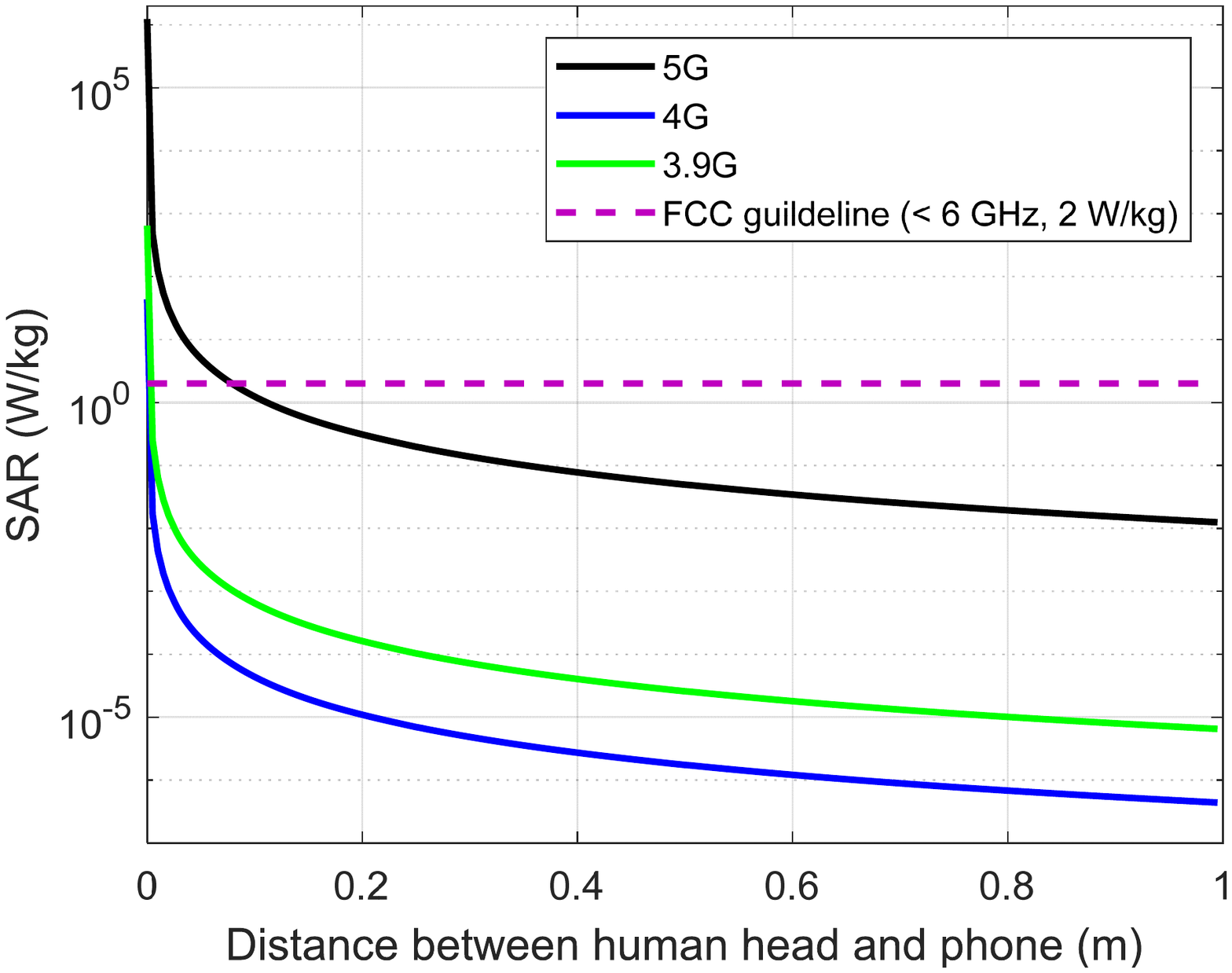}
\caption{Uplink}
\label{fig_uplink}
\end{subfigure}
\caption{Comparison of SAR among wireless technologies in downlink and uplink}
\label{fig_sar}
\end{figure*}

\vspace{0.1 in}

\section{Contributions and Current Results}
The results from our work are provided in Figure 1. We notice that 5G is marking higher SAR than the previous-generation systems mainly due to the higher carrier frequency \cite{mine5}--i.e., 28 GHz for 5G, 2 GHz for 4G, and 1.9 GHz for 3.9 GHz. The terms 5G, 4G, and 3.9G indicate the standards of Releases 14, 12, and 9 by the 3rd Generation Propect Partnership (3GPP), respectively.

Here is the reasoning of a higher SAR from a higher carrier frequency. The SAR is defined as the ratio of PD applied on the skin to the penetration depth. An EMF with a higher carrier frequency incurs a higher SAR since it can penetrate a shallower depth.

Another key reason that 5G shows a higher SAR is its small cell size--i.e., 200 m for 5G, 500 m for 4G, and 1,000 m for 3.9G. Although the EMF experiences a higher attenuation due to the higher carrier frequency, the signal power bounces back up again more frequently due to the deployment of smaller cells.

\vspace{0.1 in}

\section{Conclusions and Future Work}
Results from this project will advance the state-of-the-art solutions to the human EMF exposure problem along two directions. Existing approaches to human health-aware cellular networking usually (i) considered uplink only; and/or (ii) controls transmit power of a single user equipment (UE) only.

This work showed the following contributions that will be exploited to improve understanding. First, design of a downlink mechanism that protects human was provided. Deployment of more BSs and narrower beams in future cellular networks might increase the negative impacts of EMF on human body. Thus, although operating further than UEs from humans, BSs also need to be controlled to minimize human exposure. Second, enhancement of the uplink schemes were provided to a systematic approach to enhance the end-user performance. When an uplink transmission violates the SAR threshold, instead of reducing the UE’s transmit power, the UE is handed over to another BS with the minimum emission toward the user.

We identify the following two approaches as future work embodying the propositions of this work. The first approach is an \textit{intra-system} mechanism. Optimization and stochastic geometry will be used to assess the performance of the proposed protocols given the user information distributed in a network. Second, an \textit{inter-system} aprpoach will also be possible. The investigation will be extended to modeling use of the external networks' resource for selecting a BS with minimized human EMF exposure. The performance of the proposed protocols based on database-aided spectrum sharing will be understood via mathematical analysis and experiment on a testbed.

\vspace{0.1 in}


\end{document}